\documentclass[
    ,final            
  ]
  {aipproc}

\usepackage{amsmath,amssymb}

\layoutstyle{8x11single}

\def\eg{{\it e.g.}}

\begin{document}

\title{Construction and measurements of a vacuum-swing-adsorption
  radon-mitigation system}


\author{R.W.~Schnee}{
  address={Department of Physics, Syracuse University, Syracuse, NY 13244}
}

\author{R.~Bunker}{
  address={Department of Physics, Syracuse University, Syracuse, NY 13244}
}

\author{G.~Ghulam}{
  address={Department of Physics, Syracuse University, Syracuse, NY 13244}
}

\author{D.~Jardin}{
  address={Department of Physics, Syracuse University, Syracuse, NY 13244}
}

\author{M.~Kos}{
  address={Department of Physics, Syracuse University, Syracuse, NY 13244}
}

\author{A.S.~Tenney}{
  address={Department of Physics, Syracuse University, Syracuse, NY 13244}
}

\begin{abstract}
Long-lived alpha and beta emitters in the $^{222}$Rn decay chain on (and near) detector surfaces may be the limiting background in many experiments attempting to detect dark matter or neutrinoless double beta decay, and in screening detectors.
In order to reduce backgrounds from radon-daughter plate-out onto the wires of the BetaCage during its assembly, an ultra-low-radon cleanroom is being commissioned at Syracuse University using a vacuum-swing-adsorption radon-mitigation system.  The radon filter shows $\sim$20$\times$ reduction at its output, from $7.47\pm0.56$ to $0.37\pm0.12$\,Bq/m$^{3}$,  
and the cleanroom radon activity meets project requirements, with a lowest achieved value consistent with that of the filter, and levels consistently $<2$\,Bq/m$^{3}$.
\end{abstract}

\maketitle


\section{Introduction to radon mitigation}
\label{sec:intro}

A potentially dominant background for many rare-event searches 
or screening detectors
is from radon daughters deposited from the atmosphere
onto detector components.
Examples include $^{214}$Po for SuperNEMO~\cite{LRT2013SuperNEMO};
$^{210}$Pb for EDELWEISS~\cite{LRT2013edelweiss}, SuperCDMS~\cite{LRT2013SuperCDMS} and
the BetaCage~\cite{LRT2013bunker};
$^{210}$Po for CUORE~\cite{LRT2013cuore};
the $^{206}$Pb recoil nucleus from $^{210}$Po $\alpha$ decay for CRESST~\cite{cresst2012}, DEAP/CLEAN~\cite{deap2011surface}, and SuperCDMS~\cite{LRT2013SuperCDMS};
and neutrons from ($\alpha,n$) reactions on Teflon for 
LUX, XENON1T, and DArKSIDE.

To protect detector components, assembly within vacuum glove boxes and/or cleaning after assembly (\eg~\cite{LRT2013schneeEP}) may be effective.  However, 
vacuum glove boxes are impractical for large objects or for delicate assembly that could be jeopardized by reduced feel and range of motion.
Similarly, cleaning after assembly may be difficult and risky for complicated structures.
For these cases, providing radon-reduced air in a breathable atmosphere may be necessary .

There are two basic types of radon-mitigation systems: those with continuous flow through a single filter (typically of activated charcoal), and ``swing'' systems with flow alternating through two or more different filters.
Continuous systems (\eg~\cite{nemoLRT2006}) are designed so that
most radon decays before it exits the filter.  For an ideal column, the final radon concentration
$C_{\mathrm{final}} = C_{\mathrm{initial}} \exp(-t/t_{\mathrm{Rn}})$,
where $C_{\mathrm{initial}}$ is the concentration of the input air, $t$ is the characteristic breakthrough time of the filter, and $t_{\mathrm{Rn}} = 5.5$\,days is the Rn lifetime.
In order to have a sufficiently large breakthrough time to be effective, the carbon must be cooled.
Continuous systems are relatively simple and robust, are available commercially, and typically achieve reduction factors of $\sim1000$, to $\sim$10--30\,mBq/m$^{3}$.

In a swing system, one stops gas flow well before the breakthrough time $t$, and regenerates the first filter column while switching flow to a second column.
For an ideal column, no radon reaches the output.
Swing systems are more complicated than continuous systems (both in terms of their analysis and operation).
Vacuum-swing systems (\eg~\cite{LRT2004Pocar,PocarThesis}) can potentially provide better performance than a continuous system at lower cost.
Temperature-swing systems (\eg~\cite{LRT2010HallinRadon})
should provide the best performance, albeit at the highest cost and complexity.
Due to its potential and especially its low cost, we chose to build a vacuum-swing system in order to achieve the relatively modest radon reduction needed for construction of the BetaCage at Syracuse~\cite{LRT2013bunker}.

A vacuum-swing adsorption system takes advantage of the filtering medium's greater adsorption capacity at high pressures.
The carbon is regenerated by flowing a small fraction $f$ of filtered gas of mass flow $F$ back through the tank at low purge pressure $P_\mathrm{purge}$. 
The volume purge flow
\begin{equation}
\phi_{\mathrm{purge}} = \frac{P_\mathrm{atm}} {P_\mathrm{purge}} f F  = \frac{P_\mathrm{atm}} {P_\mathrm{purge}} f \phi_{\mathrm{feed}}  .
\end{equation}
On each cycle, the radon front is pushed back more than it moves forward if  the volume flow gain $G \equiv \phi_\mathrm{purge} / \phi_\mathrm{feed}>1$, that is if $f P_\mathrm{atm} > P_\mathrm{purge}$ .

 \begin{figure}[tb!]
\centering
\includegraphics[height=2.25in]{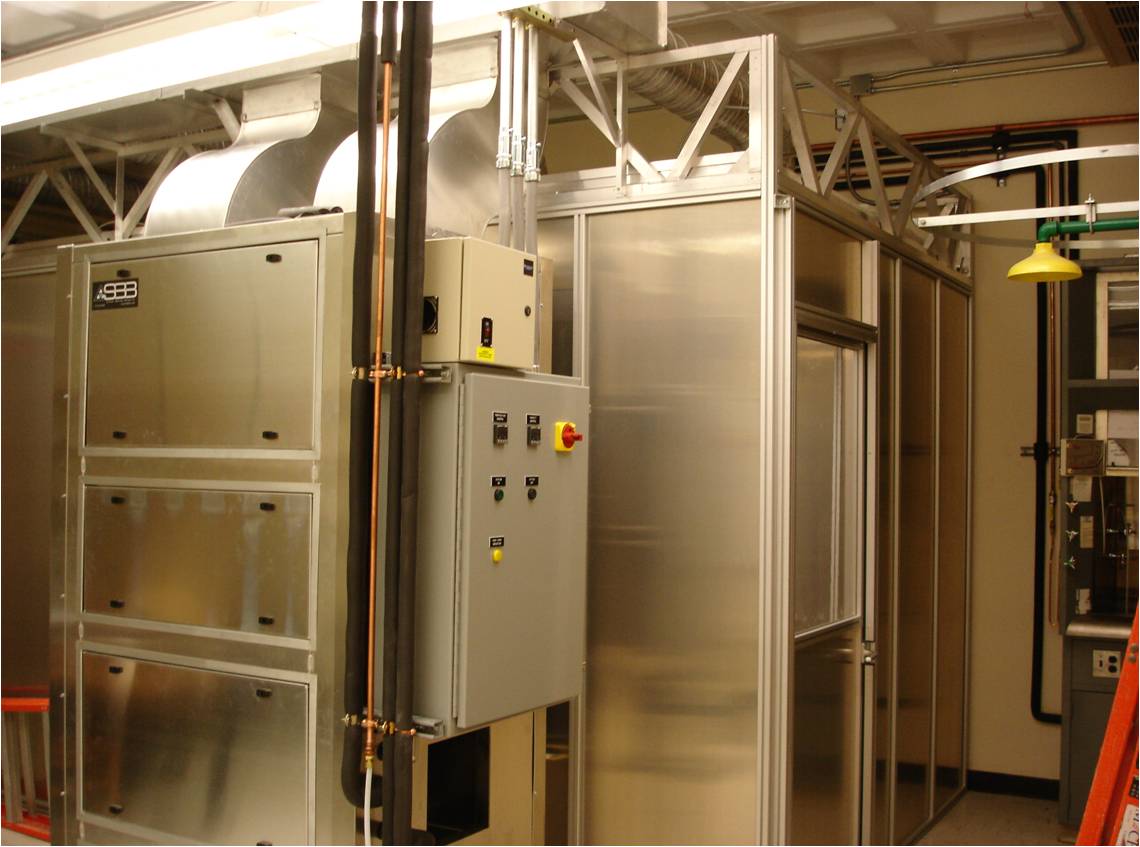}
\includegraphics[height=2.25in]{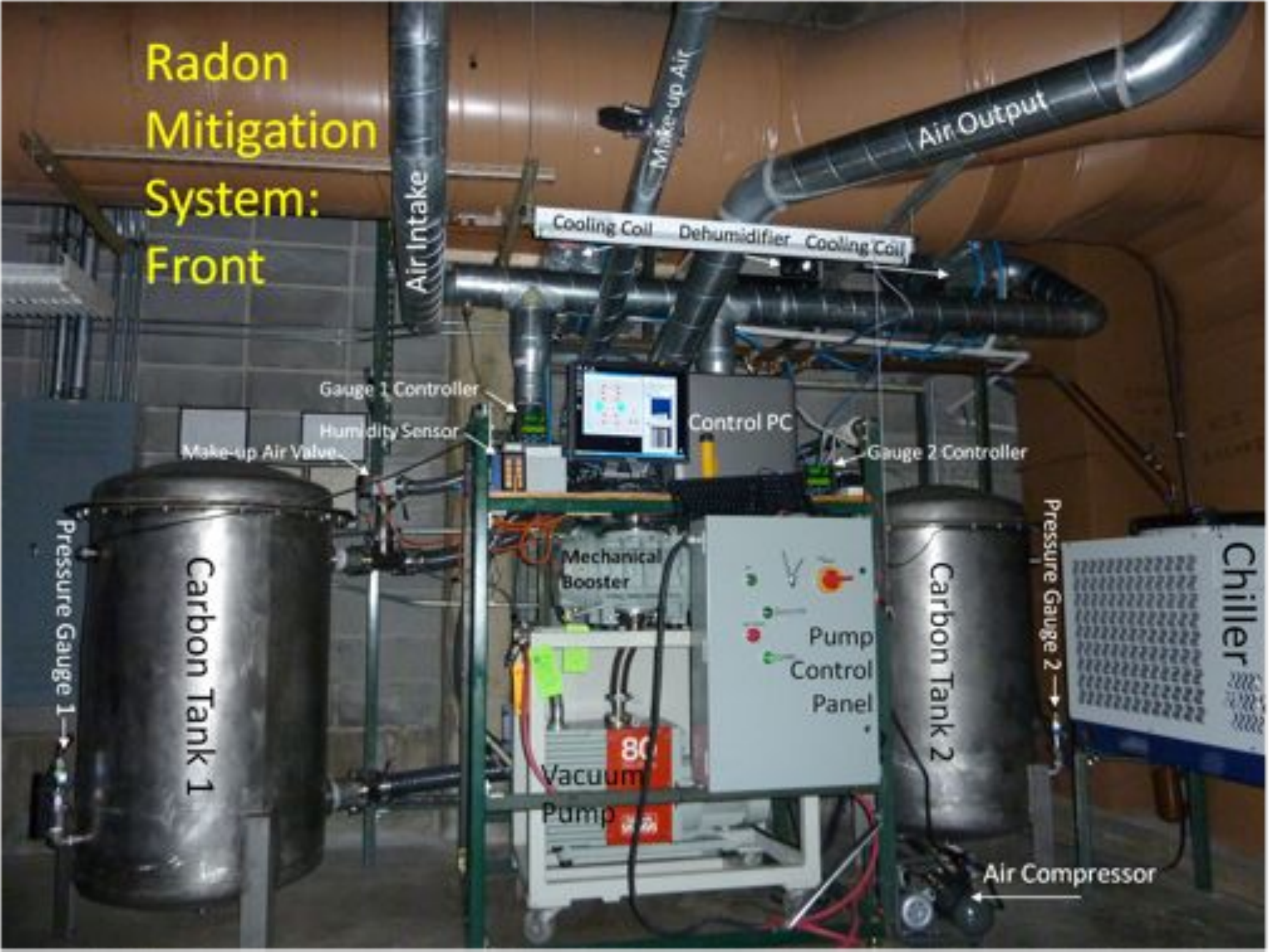}
\caption{{\it Left}: Photo of the cleanroom exterior, showing the externally housed HVAC system (before retrofitting) and the anteroom door.
{\it Right}: The radon-mitigation system.  Air ($\sim$10\,Bq/m${^3}$) enters through the top intake, then passes through a cooling coil, dehumidifier, and a second cooling coil in order to reduce the air's dew point to $<-12$\,C,  before flowing through Carbon Tank~1 or 2.  A small fraction of the air passes back through the other tank to the vacuum pump, while most of it passes through a filter (where it is monitored with a RAD7) and to the output duct that supplies air to the cleanroom one floor below.}
\label{fig:setup}
\end{figure}

\section{Design and Construction of the Syracuse radon-mitigation system}
\label{sect:system} 

In order to increase the effectiveness of filtration on removal of radon daughters and to limit radon sources, the cleanroom itself (see left panel of Fig.~\ref{fig:setup}) was designed to be as small as practically possible.  The room is 8\,ft by 12\,ft by 8\,ft high,
with a 4\,ft by 8\,ft anteroom.
To minimize emanation and permeation, it uses
all aluminum panels and extrusions, with 
thick acrylic windows, and steps were taken to ensure that the room is 
very leak tight.
The HVAC is outside the cleanroom and required retrofitting  to make it sufficiently leak tight not to be a dominant source of radon.
Aged water is used for humidification.
The room was designed for 30\,cfm low-radon makeup air, with fast HEPA filtration recirculating at a rate of one air exchange per 43 seconds.

\begin{table}[bt!]
\begin{tabular}{lrrr}
\hline
 \tablehead{1}{l}{b}{Item}
  & \tablehead{1}{r}{b}{Princeton (US\$)}
  & \tablehead{1}{r}{b}{Syracuse (US\$)}
  & \tablehead{1}{r}{b}{Comments}
   \\
\hline
tanks 						& 8k 		& 9k 		&  specweld.com   \\
charcoal 						& 6k 		& 1.5k 	& less and different carbon (see text)  \\
vacuum pumps					& 22k 	& 10k & roughing pump has lower capacity (see text) \\
valves						& 4k		& 7k		&  VAT  3" (2" for purge and make-up air) \\
dryer							& 3.5k	& 7.5k	&  Munters HC-150 and Hack Air cooling coils   \\
blower						& 1.5k	& (none)	&     \\
filter + housing					& 1.5k	& 1k		&   Clark Air ASHRAE filters at input and output \\
PC and valve control boards		& 1.5k	& 6k 		& including gauges \\
other (fittings, tubing, \itshape{etc.})			& 5k		& 5k + 8k chiller &   Pro Air Plus ACCPS015-2B  \\
\hline
total					&	53k			& 55k 	&     \\
\hline
\end{tabular}
\caption{Comparison of costs of components for 2004 Princeton~\cite{PocarThesis} and 2013 Syracuse VSA systems.}
\label{tab:a}
\end{table}

The Syracuse radon-mitigation system (see right panel of Fig.~\ref{fig:setup}) was based closely on the Princeton design~\cite{LRT2004Pocar,PocarThesis}. We tried to make some improvements focusing on ensuring that the radon reduction at the filter output was realized in the cleanroom, and we cut several corners in an effort to minimize costs.
Table~\ref{tab:a} compares the costs of components for the Princeton~\cite{PocarThesis} and Syracuse systems. Most notably, we use a roughing pump (Edwards E2M80) with significantly lower capacity at high pressures (and a significant cost savings).  We also use about 60\% as much carbon (possible due to the $\sim$2$\times$ lower airflow).  Because the carbon used in the Princeton system is no longer available, we selected 
the most similar product available (Calgon Coconut Activated Carbon Product OVC Plus 4$\times$8 mesh).
The carbon was multiply rinsed (with a final rinse in deionized water), then dried under high-flow fume hoods.
Two identical stainless-steel vacuum vessels were filled with $\sim$150\,kg each and spring-loaded in order to maintain firmly packed columns during swing operation.
As a check, we opened a tank after the first month of commissioning, finding that the carbon was still in good shape and well packed.

\begin{figure}[tb!]
\centering
\includegraphics[height=2.2in]{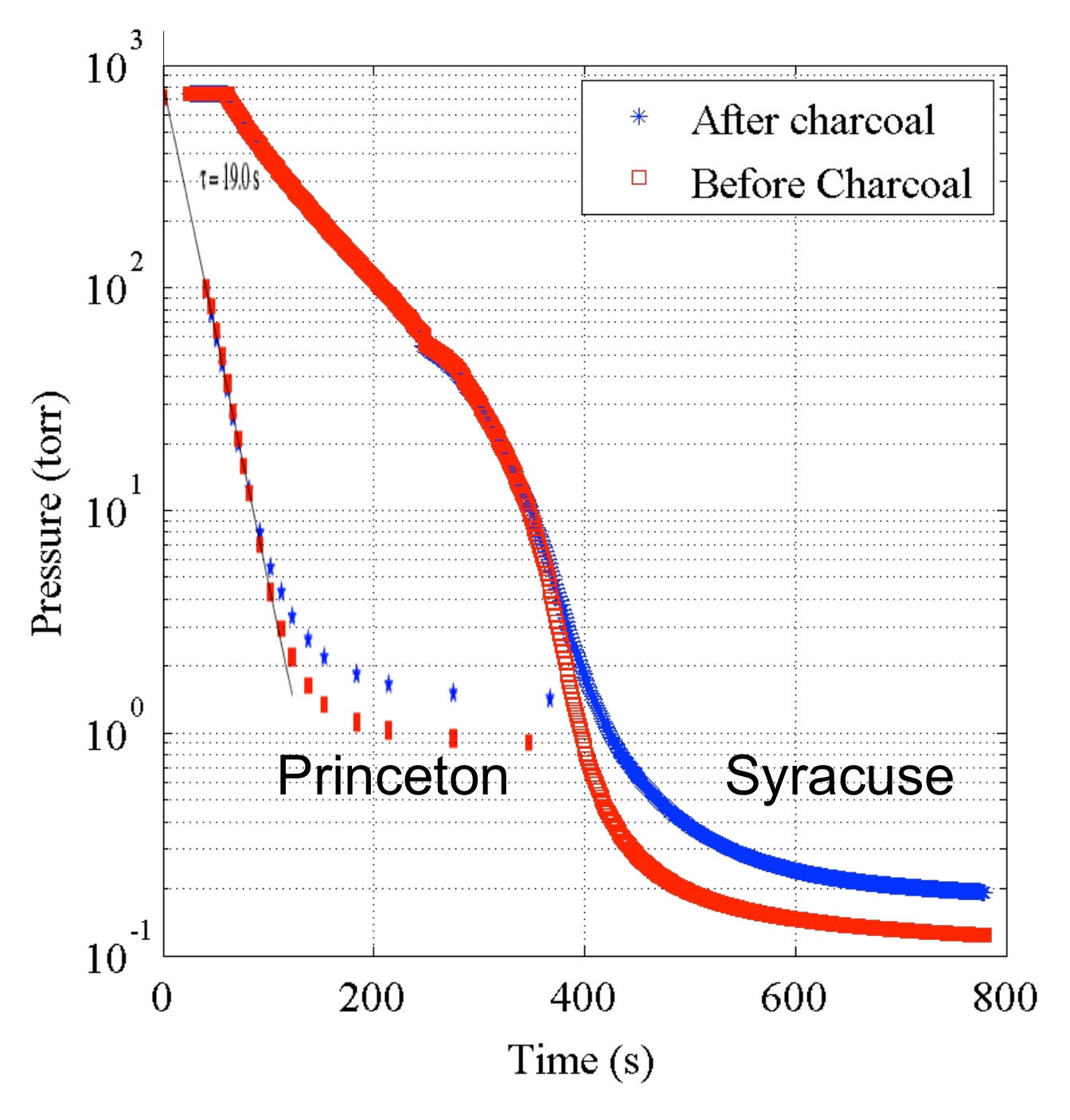}
\includegraphics[height=2.2in]{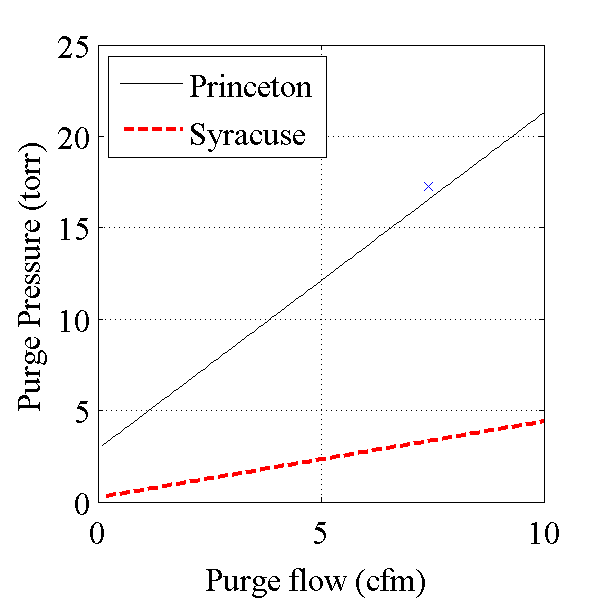}
\includegraphics[height=2.2in]{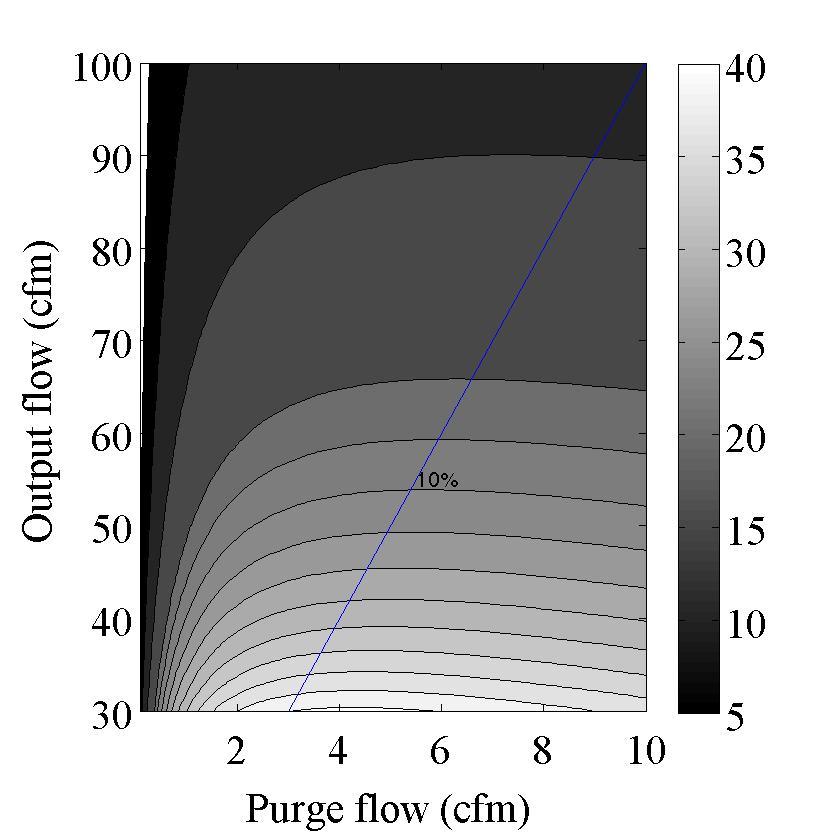}
\caption{{\it Left}: Pressure in a carbon tank on the pump side (light squares) and far side (dark stars) of the charcoal as a function of time since start of pump down, for the Princeton (small symbols) and Syracuse (large symbols) systems.  The Syracuse system takes 4\,min longer to pump down to 10\,Torr, but also achieves a lower base pressure. 
{\it Center}: Purge pressure as a function of the purge flow for the Princeton (thin solid) and Syracuse (thick dash) systems.  
{\it Right}: Predicted volume flow gain $G$ (grayscale) as a function of the output and purge flows of the Syracuse system.  Lower output flow results in higher $G$ but is less effective overcoming emanation or leaks in the cleanroom.
}
\label{fig:predictions}
\end{figure}

As shown in the left panel of Fig.~\ref{fig:predictions}, the lower capacity of the Syracuse pump at high pressures leads to a 5-min pump down to $\sim$10\,Torr (vs. Princeton $\sim$1\,min), so part of the purging cycle is inefficient.
However, the lower achieved base pressure allows the Syracuse system to operate at a lower purge pressure for the same purge flow, as shown in the center panel of Fig.~\ref{fig:predictions}, resulting
in a  high predicted  volume flow gain $G$, as shown in the right panel of Fig.~\ref{fig:predictions}.  

Although system performance is best in principle when the swing period is short compared to the breakthrough time, the finite pump-down time limits the minimum swing period in practice.  
All results described here use a swing period of 90\,min 
(although further optimization may be possible); during each 45-min half-period, one column filters air for the cleanroom while the other is evacuated for $\sim$5\,min to $<$10\,Torr, regenerated for $\sim$39\,min, and finally repressurized to 1\,atm for 1\,min (which also acts as necessary dead time for recovery of the roughing pump).  Measurements to determine the optimal output flow rate are ongoing, while the purge flow rate is typically set to $\sim$3\,cfm. 

\section{Current radon-mitigation results and status}

\begin{figure}[tb!]
\centering
\includegraphics[height=2.25in]{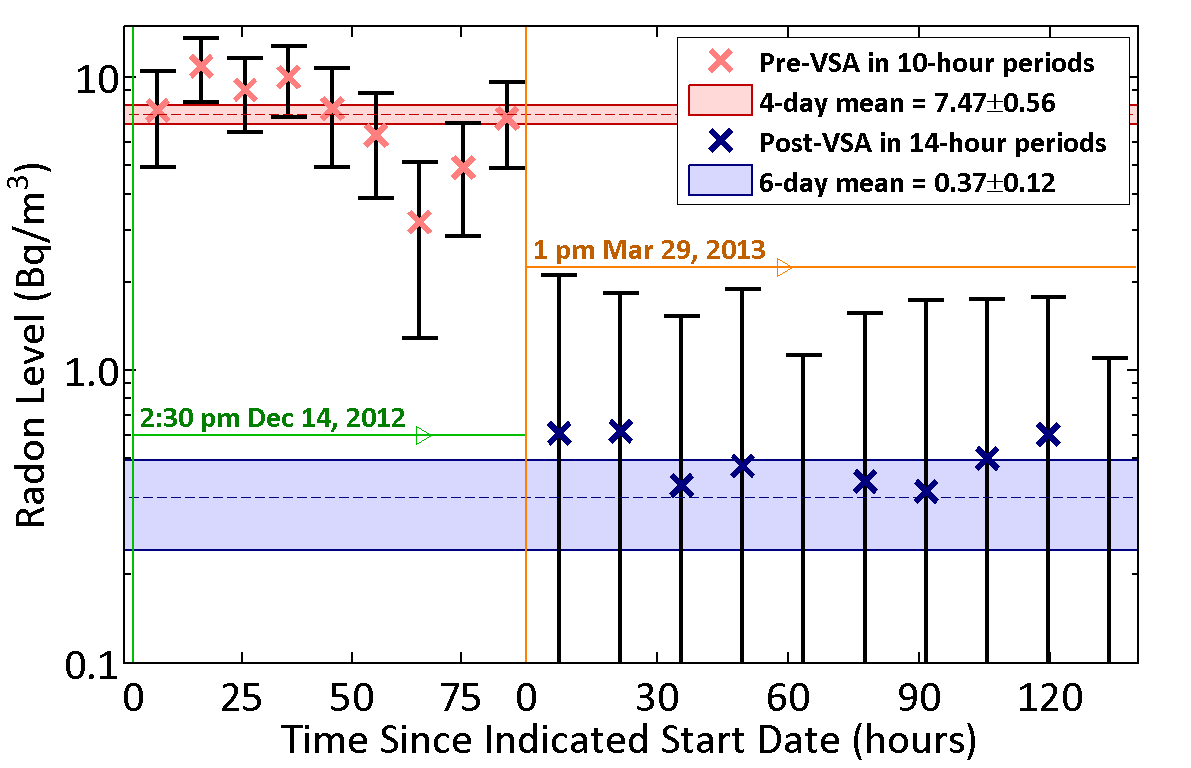}
\includegraphics[height=2.25in]{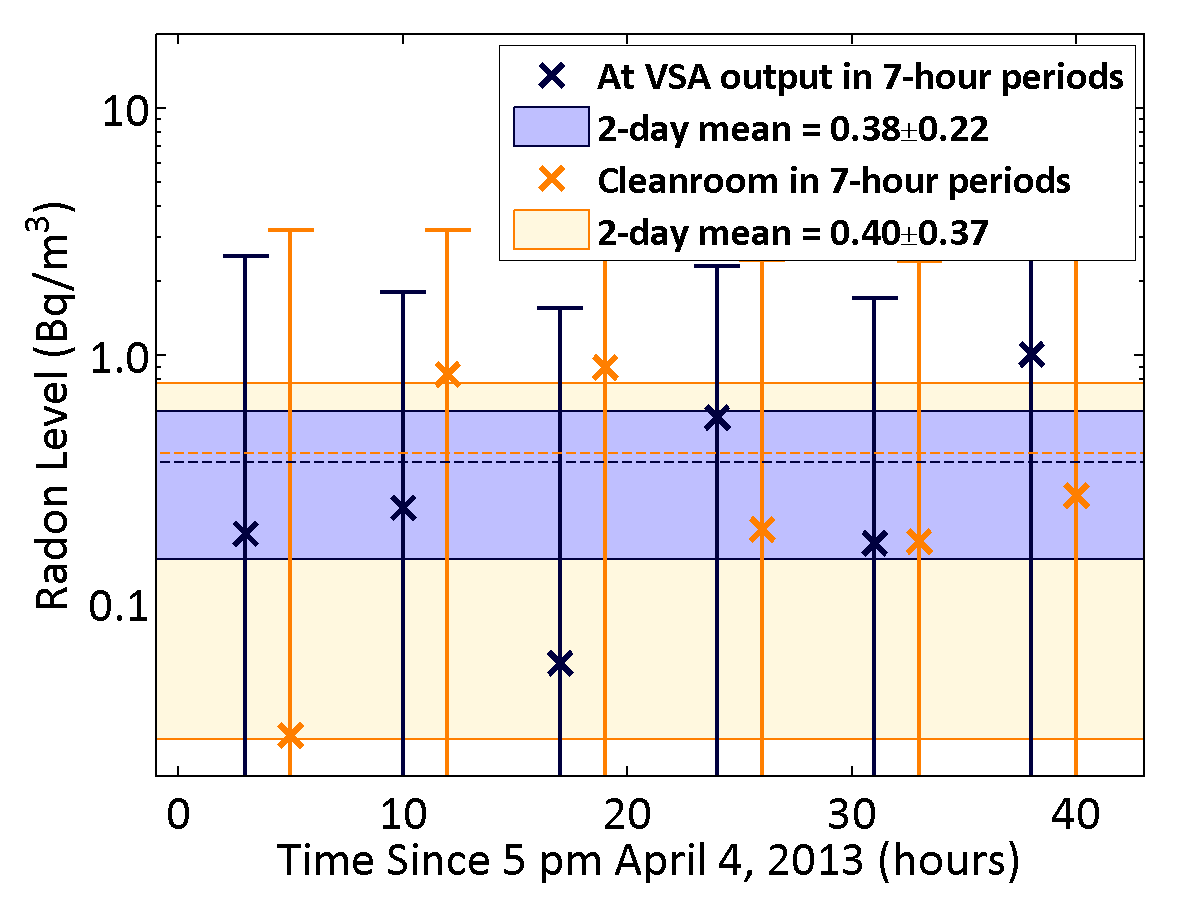}
\caption{{\it Left}: Radon activities measured for the input air (light $\times$'s and upper band) and output air (dark $\times$'s and lower band) of the VSA filter,
after subtraction of the RAD7's intrinsic background (measured with boil-off nitrogen to be $0.19\pm0.03$\,Bq/m$^{3}$ and consistent with expectation).
Average values (shaded bands) of the multi-hour periods (error bars) derived by rebinning RAD7 measurements of $^{214}$Po and  $^{218}$Po alpha decays (originally taken in 1-hour intervals) indicate $\sim$20$\times$ reduction, from 
$7.47\pm0.56$ to $0.37\pm0.12$\,Bq/m$^{3}$.
{\it Right}:  Radon activities measured at the filter output (dark $\times$'s and narrow band around lower dashed line) and within the cleanroom (light $\times$'s and wide band around higher dashed line).  Average values (shaded bands) of the 7-hour periods (error bars) indicate a cleanroom radon activity 
consistent with the level measured at the filter's output.
Uncertainties on the cleanroom activity are relatively 
large due to
measurement with an older RAD7 with a larger intrinsic background (of $0.67\pm0.10$\,Bq/m$^{3}$).
Note that the RAD7 uncertainties quoted here are known to be overly conservative.
}
\label{fig:results}
\end{figure}

The radon-mitigation system was first turned on in December 2012.
First results, shown in the left panel of Fig.~\ref{fig:results}, indicated a radon reduction of $\sim$20$\times$, to $0.4\pm0.1$\,Bq/m$^{3}$.
Assuming a dynamic adsorption coefficient of 3\,m$^{3}$/kg (STP), which is on the low end of the range for typical activated carbon, the predicted breakthrough time for each column is 4.2\,h for a
volume flow rate $\phi_{\mathrm{feed}} = 60$\,cfm.  
By simply switching from the swing cycle to flowing air continuously through a single tank, a lower limit on the actual breakthrough time of about two hours was measured.

First measurement of the radon level in the clean room, with the cleanroom air circulation off, was performed in April 2013, as shown in Fig.~\ref{fig:results}.  The measured radon activity of 0.4\,Bq/m$^{3}$ was the same as that measured at the input duct to the cleanroom and indicates that emanation and leaks in the cleanroom itself are minor.  Subsequent measurements with the air circulation on indicated significant leaks in the air circulation path which had to be fixed.  Operation of the radon filter since this time with various configurations has led to a range of radon output levels (some consistent with zero), averaging to $0.33\pm0.13$\,Bq/m$^{3}$, while 
levels in the cleanroom have been consistently $<2$\,Bq/m$^{3}$.  


The 0.4\,Bq/m$^{3}$ output level of the radon filter is slightly worse than that of the Princeton Borexino system on which it was based.  Optimization of the feed and purge flows and cycle times may yield improved performance.
Moreover, we expect planned improvements to the leak tightness of the filter's output duct will bring the cleanroom radon concentration (with circulation on) in line with the filter's output.
Even so, the cleanroom radon concentration already achieved is better than the level required  for construction of the BetaCage~\cite{LRT2013bunker}.  The concentration is lower than that achieved for the 
Princeton system (by $\sim4\times$ with the circulation off), likely due to reduced emanation from the cleanroom itself.  
The results indicate that the VSA technique remains a viable lower-cost alternative to radon mitigation using cooled carbon filters.


\begin{theacknowledgments}
This work was supported in part by the 
National Science Foundation (Grant No.\ PHY-0855525).  
The authors thank T.~Shutt, A.~Hallin, A.~Pocar, W.~Rau, and V. Guiseppe for useful discussions and advice.

\end{theacknowledgments}



\bibliographystyle{aipproc}   

\bibliography{schnee}

\begin{thebibliography}{12}
\expandafter\ifx\csname natexlab\endcsname\relax\def\natexlab#1{#1}\fi
\providecommand{\enquote}[1]{``#1''}
\expandafter\ifx\csname url\endcsname\relax
  \def\url#1{\texttt{#1}}\fi
\expandafter\ifx\csname urlprefix\endcsname\relax\def\urlprefix{URL }\fi
\providecommand{\eprint}[2][]{\url{#2}}

\bibitem[{Mott}(2013)]{LRT2013SuperNEMO}
J.~{Mott}, \enquote{{Low-background tracker development for the SuperNEMO
  experiment},} in \emph{Topical Workshop on Low Radioactivity Techniques: LRT
  2013}, edited by L.~{Miramonti}, and L.~{Pandola}, 2013, vol. 1549 of
  \emph{American Institute of Physics Conference Series}, pp. 152--155.

\bibitem[{Navick}(2013)]{LRT2013edelweiss}
X.-F. {Navick}, \enquote{{Background suppression in the EDELWEISS-III
  experiment},} in \emph{Topical Workshop on Low Radioactivity Techniques: LRT
  2013}, edited by L.~{Miramonti}, and L.~{Pandola}, 2013, vol. 1549 of
  \emph{American Institute of Physics Conference Series}, pp. 148--151.

\bibitem[{Cooley}(2013)]{LRT2013SuperCDMS}
J.~{Cooley}, \enquote{{Background Considerations for SuperCDMS},} in
  \emph{Topical Workshop on Low Radioactivity Techniques: LRT 2013}, edited by
  L.~{Miramonti}, and L.~{Pandola}, 2013, vol. 1549 of \emph{American Institute
  of Physics Conference Series}, pp. 223--226.

\bibitem[{Bunker} et~al.(2013)]{LRT2013bunker}
R.~{Bunker}, M.~A. {Bowles}, M.~{Kos}, R.~W. {Schnee}, B.~{Wang}, Z.~{Ahmed},
  S.~R. {Golwala}, R.~H. {Nelson}, A.~{Rider}, A.~{Zahn}, and D.~R. {Grant},
  \enquote{{The BetaCage, an Ultra-sensitive Screener for Surface
  Contamination},} in \emph{Topical Workshop on Low Radioactivity Techniques:
  LRT 2013}, edited by L.~{Miramonti}, and L.~{Pandola}, 2013, vol. 1549 of
  \emph{American Institute of Physics Conference Series}, pp. 132--135.

\bibitem[{Pattavina}(2013)]{LRT2013cuore}
L.~{Pattavina}, \enquote{{Radon induced surface contaminations in low
  background experiments},} in \emph{Topical Workshop on Low Radioactivity
  Techniques: LRT 2013}, edited by L.~{Miramonti}, and L.~{Pandola}, 2013, vol.
  1549 of \emph{American Institute of Physics Conference Series}, pp. 82--85.

\bibitem[{Angloher} et~al.(2012)]{cresst2012}
G.~{Angloher}, M.~{Bauer}, I.~{Bavykina}, A.~{Bento}, C.~{Bucci},
  C.~{Ciemniak}, G.~{Deuter}, F.~{von Feilitzsch}, D.~{Hauff}, P.~{Huff},
  C.~{Isaila}, J.~{Jochum}, M.~{Kiefer}, M.~{Kimmerle}, J.-C. {Lanfranchi},
  F.~{Petricca}, S.~{Pfister}, W.~{Potzel}, F.~{Pr{\"o}bst}, F.~{Reindl},
  S.~{Roth}, K.~{Rottler}, C.~{Sailer}, K.~{Sch{\"a}ffner}, J.~{Schmaler},
  S.~{Scholl}, W.~{Seidel}, M.~v. {Sivers}, L.~{Stodolsky}, C.~{Strandhagen},
  R.~{Strau{\ss}}, A.~{Tanzke}, I.~{Usherov}, S.~{Wawoczny}, M.~{Willers}, and
  A.~{Z{\"o}ller}, \emph{European Physical Journal C} \textbf{72}, 1971 (2012),
  \eprint{1109.0702}.

\bibitem[{Cai} et~al.(2011)]{deap2011surface}
B.~{Cai}, M.~{Boulay}, B.~{Cleveland}, and T.~{Pollmann}, \enquote{{Surface
  backgrounds in the DEAP-3600 dark matter experiment},} in \emph{American
  Institute of Physics Conference Series}, edited by R.~{Ford}, 2011, vol. 1338
  of \emph{American Institute of Physics Conference Series}, pp. 137--146.

\bibitem[{Schnee} et~al.(2013)]{LRT2013schneeEP}
R.~W. {Schnee}, M.~A. {Bowles}, R.~{Bunker}, K.~{McCabe}, J.~{White},
  P.~{Cushman}, M.~{Pepin}, and V.~{Guiseppe}, \enquote{{Removal of long-lived
  222Rn daughters by electropolishing thin layers of stainless steel},} in
  \emph{Topical Workshop on Low Radioactivity Techniques: LRT 2013}, edited by
  L.~{Miramonti}, and L.~{Pandola}, 2013, vol. 1549 of \emph{American Institute
  of Physics Conference Series}, pp. 128--131.

\bibitem[{Nachab}(2007)]{nemoLRT2006}
A.~{Nachab}, \enquote{{Radon reduction and radon monitoring in the NEMO
  experiment},} in \emph{AIP Conf. Proc. 897: Topical Workshop on Low
  Radioactivity Techniques: LRT 2006}, edited by P.~{Loaiza}, American
  Institute of Physics, Melville, NY, 2007, pp. 35--39.

\bibitem[{Pocar}(2005)]{LRT2004Pocar}
A.~{Pocar}, \enquote{{Low background techniques for the Borexino nylon
  vessels},} in \emph{Topical Workshop on Low Radioactivity Techniques: LRT
  2004.}, edited by B.~{Cleveland}, R.~{Ford}, and M.~{Chen}, 2005, vol. 785 of
  \emph{American Institute of Physics Conference Series}, pp. 153--162,
  \eprint{arXiv:physics/0503243}.

\bibitem[{Pocar}(2003)]{PocarThesis}
A.~{Pocar}, \emph{{Low Background Techniques and Experimental Challenges for
  Borexino and its Nylon Vessels}}, Ph.D. thesis, Princeton University (2003).

\bibitem[{Grant} et~al.(2011)]{LRT2010HallinRadon}
D.~{Grant}, A.~{Hallin}, S.~{Hanchurak}, C.~{Krauss}, S.~{Liu}, and R.~{Soluk},
  \enquote{{Low Radon Cleanroom at the University of Alberta},} in
  \emph{American Institute of Physics Conference Series}, edited by R.~{Ford},
  2011, vol. 1338 of \emph{American Institute of Physics Conference Series},
  pp. 161--163.

\end{thebibliography}

 \IfFileExists{\jobname.bbl}{}
 {\typeout{}
  \typeout{******************************************}
  \typeout{** Please run "bibtex \jobname" to obtain}
  \typeout{** the bibliography and then re-run LaTeX}
  \typeout{** twice to fix the references!}
  \typeout{******************************************}
  \typeout{}
 }

\end{document}